\documentclass[twocolumn, twocolappendix]{aastex631}

\shorttitle{Implosion Triggered SF}
\shortauthors{Romano et al.}
\begin{document}

\title{Star Formation by Supernova Implosion}

\correspondingauthor{Leonard E. C. Romano}
\email{lromano@usm.lmu.de}

\author[0000-0001-8404-3507]{Leonard E. C. Romano}
\affiliation{Universitäts-Sternwarte, Fakultät für Physik, Ludwig-Maximilians-Universität München, Scheinerstr. 1, D-81679 München, Germany}
\affiliation{Max-Planck-Institut für extraterrestrische Physik, Giessenbachstr. 1, D-85741 Garching, Germany}
\affiliation{Excellence Cluster ORIGINS, Boltzmannstr. 2, D-85748 Garching, Germany}

\author{Andreas Burkert}
\affiliation{Universitäts-Sternwarte, Fakultät für Physik, Ludwig-Maximilians-Universität München, Scheinerstr. 1, D-81679 München, Germany}
\affiliation{Max-Planck-Institut für extraterrestrische Physik, Giessenbachstr. 1, D-85741 Garching, Germany}
\affiliation{Excellence Cluster ORIGINS, Boltzmannstr. 2, D-85748 Garching, Germany}

\author[0000-0002-8759-941X]{Manuel Behrendt}
\affiliation{Universitäts-Sternwarte, Fakultät für Physik, Ludwig-Maximilians-Universität München, Scheinerstr. 1, D-81679 München, Germany}
\affiliation{Max-Planck-Institut für extraterrestrische Physik, Giessenbachstr. 1, D-85741 Garching, Germany}



\begin{abstract}
Recent hydrodynamical simulations of the late stages of supernova remnant (SNR) evolution have revealed that as they merge with the ambient medium, SNRs implode, leading to the formation of dense clouds in their center.
While being highly chemically enriched by their host SNR, these clouds appear to have similar properties as giant molecular clouds, which are believed to be the main site of star formation.
Here, we develop a simple model, in order to estimate the efficiency of the star formation that might be triggered by the implosion of SNRs.
We separately consider two cases, cyclic star formation, maintained by the episodic driving of feedback from new generations of stars; 
and a single burst of star formation, triggered by a single explosion.
We find that in the cyclic case, star formation is inefficient, with implosion-triggered star-formation contributing a few percent of the observed star-formation efficiency per free-fall timescale.
In the single-burst case, higher star-formation efficiencies can be obtained.
However, while the implosion-triggered process might not contribute much to the overall star-formation, due to the high chemical enrichment of the birth clouds, it can explain the formation of a significant fraction of metal-rich stars.
\end{abstract}



\section{Introduction} \label{sec:intro}

Star formation is easy. Giant Molecular Clouds (GMCs) with masses $\sim 10^4\, -\,10^7\,\text{M}_{\odot}$ \citep{1997ApJ...476..166W, 2017ApJ...834...57M} fragment, producing gravitationally unstable cores that collapse \citep[e.g.][]{1902RSPTA.199....1J, 2007ARA&A..45..565M, 2020SSRv..216...62R} and form stars.
Yet, a closer look reveals that our understanding of the processes linking these stages of star formation and even the formation of the GMCs themselves is far from being predictive \citep[see e.g.][]{2013ApJ...778..133L, 2020MNRAS.493.2872C}. 
Indeed, the processes underlying star formation remain to be one of the biggest puzzles in the current framework of galaxy formation and evolution.

One of these processes, which could explain the formation of star-forming GMCs is so-called \textit{triggered star-formation}\citep{1998ASPC..148..150E, 2011EAS....51...45E}. The underlying idea being that shocks, e.g. due to Supernovae (SNe) or other feedback processes, could compress the ambient gas and drive it towards gravitational instability (GI). 

Observational evidence suggests that stars form spatially and temporally correlated \citep{2015A&A...584A..36G, 2017A&A...608A.102G, 2018A&A...619A.120K, 2022Natur.601..334Z, 2023A&A...678A..71R, 2023ApJ...953..145V}, pointing towards triggered star-formation scenarios. Meanwhile, many of the proposed triggers have been confirmed in numerical simulations \citep{2022MNRAS.509..954D, 2023MNRAS.521.5712H, 2024MNRAS.52710077H}.

Recently, \citet{2024arXiv240205796R} have found that besides the classical triggering by shock-compression, SNe will eventually implode due to the pressure of the interstellar medium (ISM) and form a dense, highly chemically enriched cloud in their center, which might collapse and form stars.
In this letter, we leverage their results to give an estimate of the amount of star formation that can be expected from this process.
\newpage

\section{Methods} \label{sec:methods}

\begin{figure}
\centering
\includegraphics[width=0.5\textwidth]{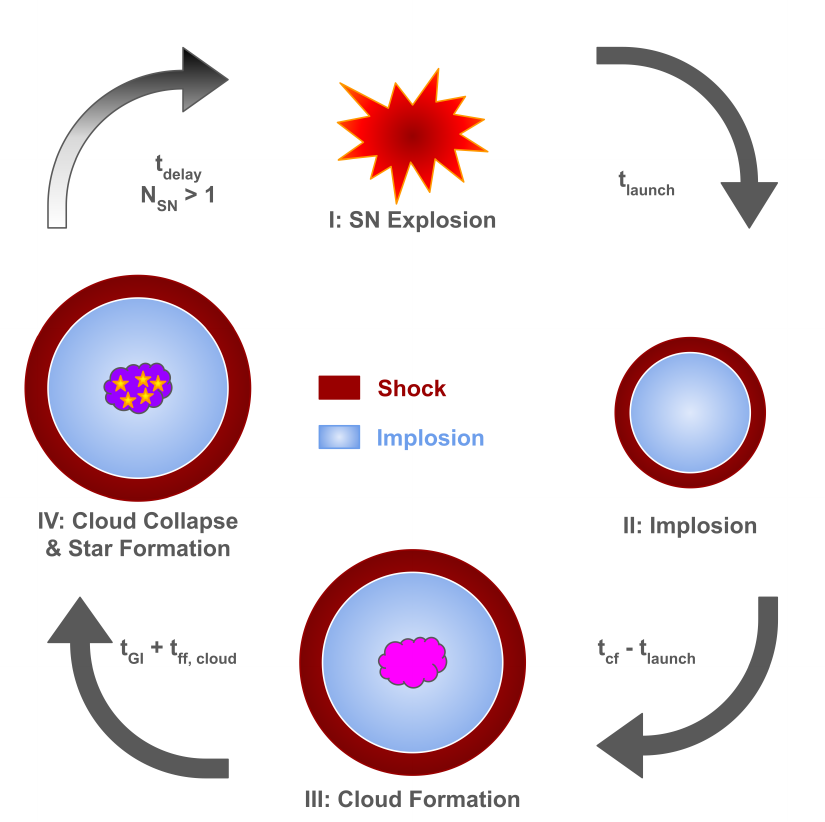}
\caption{A schematic overview of SN-implosion-triggered star-formation. In this process, an explosion from a central source drives a shock (I), which eventually implodes (II) and leads to the formation of a central cloud (III). The central cloud grows and eventually becomes gravitationally unstable, leading to star formation (IV). Depending on whether there are newly-formed, massive stars, this cycle can repeat. The arrows connecting the different stages are decorated with the corresponding transition timescales defined in equations \ref{eq:t_cycle}, \ref{eq:t_cf}, \ref{eq:t_GI} and \ref{eq:t_ff,cl}. The cloud formation time $t_{\text{cf}}$ is measured from the time of the explosion so the time between implosion and cloud formation is $t_{\text{cf}} - t_{\text{launch}}$. }\label{fig:schematic}
\end{figure}

A common way to quantify the star-formation efficiency is the so-called \textit{star-formation efficiency per free-fall time} $\epsilon_{\text{ff}}
$. It is defined as the ratio of the star-formation timescale and the free-fall timescale \citep[see e.g.][for a review]{2024arXiv240319843S}:
\begin{equation}
    \epsilon_{\text{ff}} = \frac{t_{\text{ff}}}{t_{\text{sf}}}\,.
\end{equation}
The average free-fall timescale in the ISM prior to the explosion
\begin{equation}\label{eq:t_ff}
    t_{\text{ff}} = \sqrt{ \frac{3 \pi}{32 G\rho}} \sim 44.9 \, n_{0}^{-0.5} \, \text{Myr}\,,
\end{equation}
where $n_0$ is in units of cm$^{-3}$. $t_{\text{ff}}$ depends only on the density $\rho = \mu\, m_{\text{H}} \, n_{0}$ of the ambient medium, where $\mu = 1.4$ is the mean atomic weight and $m_{\text{H}}$ is the mass of a hydrogen atom. On the other hand, the star-formation timescale
\begin{equation}
    t_{\text{sf}} = \left(\frac{1}{\rho} \left.\frac{\text{d}\rho}{\text{d}t}\right|_{\text{sf}}\right)^{-1}\,,
\end{equation}
depends on the details of the processes that convert gas into stars.

Here we derive an expression for the timescale of SN-implosion-triggered star-formation. We consider two cases. In both cases a central stellar population drives a shock which will implode due to the pressure of the ISM, leading to the formation of a central cloud \citep{2024arXiv240205796R}. Once the cloud becomes gravitationally unstable it collapses and forms new stars.
In the first, \textit{cyclic} case, we consider the long term average of stellar populations that can produce enough massive stars to keep maintaining a continuous cycle. In the second, \textit{single-burst} case we consider a stellar population or a single star driving a single iteration of SN-implosion-triggered star-formation, which may then either continue on indefinitely as in the cyclic case, or cease due to the lack of newly formed massive stars. The main difference between the two scenarios is that in the former, the number of SN explosions per cycle is fixed by the expected stellar mass that is formed by each cycle, while in the latter it is kept as a free parameter. A schematic overview of such a cycle is shown in Figure \ref{fig:schematic}.

We utilize the results of recent simulations by \citet{2024arXiv240205796R}. Their results include timescales for various stages of SN Remnant (SNR) evolution and cloud formation as well as information about various cloud properties as a function of time after its formation (see e.g. their figures 7 and 8). 

Besides these results, we need to make the following simplifying assumptions:
\begin{enumerate}
    \item Stellar populations follow the \textit{canonical Initial Mass Function} (IMF) $\xi_{\text{star}}\left(m\right)$ from \citet{2021arXiv211210788K}.
    \item Implosion-triggered star-formation can only be sustained if it leads to the formation of at least one massive ($m_{\star} > 8 \, \text{M}_{\odot}$) star, which can trigger another SN implosion.
    \item The ambient medium is initially at rest, with a uniform density of $n_{\text{H}} = n_0$ and returns to this state by the end of each cycle.
    \item We neglect the role of early feedback (i.e. stellar winds and ionizing stellar radiation) and assume that all stars in the stellar population explode simultaneously in a common location, injecting a total energy of $E_{\text{SN}} = 10^{51}\,\text{erg}$ per SN into the ambient medium.
    \item We neglect all other sources of star formation and stellar feedback.
\end{enumerate}

The star-formation-rate per unit volume for this process can be written as
\begin{equation}
    \left.\frac{\text{d}\rho}{\text{d}t}\right|_{\text{sf}} = \frac{M_{\star\text{, formed}}}{t_{\text{cycle}} V_{\text{exp}}}\,,
\end{equation}
where the stellar mass formed per cycle $M_{\star\text{, formed}} = \epsilon_{\star} \, M_{\text{cloud}}$ is parameterized as a fraction $\epsilon_{\star}$ of the mass of the central cloud prior to collapse, $V_{\text{exp}} = 4\pi/3 \, R_{\text{exp}}^3$ is the volume traced by the shock during its expansion, and 
\begin{equation}\label{eq:t_cycle}
    t_{\text{cycle}} = t_{\text{delay}} + t_{\text{cf}} + t_{\text{GI}} + t_{\text{ff, cloud}}\,,
\end{equation}
is the time it takes to complete one cycle.

The fraction of cloud mass that is converted into stars $\epsilon_{\star}$ has been constrained both from observations \citep{2016ApJ...833..229L, 2020MNRAS.493.2872C} and simulations \citep{2022MNRAS.512..216G, 2024MNRAS.527.6732F}. It shows large variation with a typical value of $\epsilon_{\star} \sim 10 \, \%$. Due to its large uncertainties, we treat this number as a model parameter, with a fiducial value of $\epsilon_{\star}^{\text{fid}} = 10 \, \%$.

In summary the resulting free-fall efficiency, associated with SN-implosion-triggered star-formation can be compactly written as
\begin{equation}
    \epsilon_{\text{ff}} = \epsilon_{\star} \frac{M_{\text{cloud}}}{M_{\text{exp}}} \frac{t_{\text{ff}}}{t_{\text{cycle}}} \,
\end{equation}
where $M_{\text{exp}} = \mu m_{\text{H}} n_{0} V_{\text{exp}}$ is the total mass swept up by the SNR during it's expansion, and
$t_{\text{ff}}$, $t_{\text{cycle}}$ and $M_{\text{cloud}}$ are given by eqs. \ref{eq:t_ff}, \ref{eq:t_cycle} and \ref{eq:cloud_mass}, respectively.

There is some delay before the massive stars in a newly formed population explode. This time delay has been computed by \citet{2017AJ....153...85S} under the single stellar population approximation and is typically in the range of $t_{\text{delay}} \sim 2.2 \, - \, 4.6 \, \text{Myr}$. We treat this timescale as a model parameter with a fiducial value of $t_{\text{delay}}^{\text{fid}} \sim 3 \, \text{Myr}$. However, we note that for stellar populations with exclusively long-lived massive stars, the time delay can be over an order of magnitude longer.

For $n_0 \gtrsim 1$ the cloud-formation timescale \citep{2024arXiv240205796R}
\begin{equation}\label{eq:t_cf}
    t_{\text{cf}} \sim 4 \, n_{0}^{-0.15}\, N_{\text{SN}}^{0.3}\, \text{Myr}\,,
\end{equation}
where $N_{\text{SN}}$ is the number of SNe exploding simultaneously.

After its formation the central cloud evolves and its mass grows roughly like
\begin{equation}
    M_{\text{cloud}} \sim 20 \, \left(\frac{t}{1 \, \text{Myr}}\right)^{2.6} \, \text{M}_{\odot} \,,
\end{equation}
and its virial parameter drops 
\begin{equation}
    \alpha_{\text{vir}} \sim 6000 \, n_{0}^{-1} \, \left(\frac{t}{1 \, \text{Myr}}\right)^{-1.75} \, \text{M}_{\odot} \,.
\end{equation}

The cloud collapses when it becomes gravitationally unstable, i.e. $\alpha_{\text{vir}} \left(t_{\text{GI}}\right) \sim 1$, which leads to
\begin{eqnarray}\label{eq:t_GI}
    t_{\text{GI}} \sim 144 \, n_{0}^{-0.57}\, \text{Myr} ~,\\
    M_{\text{cloud}}\left(t_{\text{GI}}\right) \sim 8 \times 10^{6}\, n_{0}^{-1.49}\, \text{M}_{\odot} \, ~. \label{eq:cloud_mass}
\end{eqnarray}

After the onset of collapse, the cloud is expected to collapse on a free-fall timescale, which depends on the density of the cloud at $t_{\text{GI}}$. 

We find that the overdensity of the cloud relative to the ambient medium does not follow a simple pattern, but its overall magnitude oscillates in time, with typical values in the range of $\chi \sim 2 - 4$ and no clear correlation with the simulation parameters. We thus treat the overdensity as a model parameter with a fiducial value of $\chi^{\text{fid}} = 3$, which is justified by our results in section \ref{sec:results} that show that the value of $\chi$ only has a marginal effect on the star-formation efficiency. The collapse timescale of the cloud is thus 
\begin{equation}\label{eq:t_ff,cl}
    t_{\text{ff, cloud}} = \chi^{-0.5} \, t_{\text{ff}} \sim 44.9 \, \chi^{-0.5} \, n_{0}^{-0.5}\, \text{Myr}\,.
\end{equation}

In the meantime the blastwave expansion has affected a spherical region of radius $R_{\text{exp}}$. Assuming that the blastwave expands as a momentum-conserving snowplow, i.e. $R \propto t^{1/4}$, we estimate the size of the affected region as
\begin{equation}
    R_{\text{exp}} \sim R_{\text{launch}} \, \left(\frac{t_{\text{cf}} + t_{\text{GI}} + t_{\text{ff, cloud}}}{t_{\text{launch}}}\right)^{1/4}\,,
\end{equation}
where $R_{\text{launch}} \sim 51.3\,n_{0}^{-0.27}\, N_{\text{SN}}^{0.3}\,\text{pc}$ and $t_{\text{launch}} \sim 0.98\, n_{0}^{-0.11}\, N_{\text{SN}}^{0.27}\, \text{Myr}
$ \citep{2024arXiv240205796R}.

The scaling of the cloud mass eq. \ref{eq:cloud_mass} implies that above a certain density $n_{\text{max}}$, the cloud mass becomes so low that it becomes unlikely to form massive stars, which could trigger the next cycle of star formation. 

By normalizing the IMF to the stellar mass formed from the collapsing cloud, we find that the number of massive stars is
\begin{equation}\label{eq:N_SN}
    N_{\text{SN}} = \int_{8 \,\text{M}_{\odot}}^{\infty} \xi_{\text{star}}\left(m\right)\text{d}m \sim 1.08 \, \frac{\epsilon_{\star} M_{\text{cloud}}}{100 \, \text{M}_{\odot}}\,,
\end{equation}
which in combination with eq. \ref{eq:cloud_mass} implies that the maximum density at which one can expect the formation of at least one massive star is
\begin{equation}\label{eq:n_max}
    n_{\text{max}} \sim 1070 \, \epsilon_{\star}^{0.67} \, \text{cm}^{-3}\,.
\end{equation}

A further constraint can be imposed by considering that the dynamics may be dramatically altered due to vertical stratification if the shock breaks out 
of the galactic ISM before the implosion is launched \citep[see e.g.][]{1990ApJ...354..513K}. Indeed, for the lowest density considered here the SNR would have expanded to a radius of $R_{\text{launch}} \sim 780 \,\text{pc}$ before imploding and thus if located in an ISM resembling the solar neighborhood with a scale height of $150 \, \text{pc}$ \citep{2015ApJ...814...13M} would have broken out before it could have imploded.

For an isothermal, single-component disk in vertical hydrostatic equilibrium the vertical density profile is $\propto \text{sech}^{2}\left(z/z_{0}\right)$ with scale height \citep{2015MNRAS.448.1007B}
\begin{equation}
    z_{0} = \frac{\sigma}{\sqrt{2\pi G \rho_{\text{mp}}}} \sim 338 \,\sigma_{10} n_{0}^{-0.5}\, \text{pc} ~,
\end{equation}
with velocity dispersion $\sigma = 10 \, \sigma_{10} \, \text{km/s}$ and midplane density $\rho_{\text{mp}}$.
By assuming that the SNe explode in the midplane and requiring that the implosion should be launched before shock break-out, one obtains the constraint
\begin{equation}\label{eq:z_sc_constraint}
    n_{\text{H}} < 3.6 \times 10^{3} \, \sigma_{10}^{4.35} N_{\text{SN}}^{-1.3}\, \text{cm}^{-3} ~.
\end{equation}
For the cyclic case $N_{\text{SN}}$ is given by eq. \ref{eq:N_SN} and thus this translates into a lower limit for the density
\begin{equation}\label{eq:n_lower_limit}
    n_{\text{min}} = 1.1 \times 10^{3}\, \sigma_{10}^{-4.6}\epsilon_{\star}^{1.38}\,  \text{cm}^{3} ~,
\end{equation}
while in the single burst case eq. \ref{eq:z_sc_constraint} provides an upper limit for the density, due to the different scaling with density of $R_{\text{launch}}$ and $z_{0}$.

In summary the density in the cyclic case is constrained to be in the range
\begin{equation}
    \text{max}\left(1.1 \times 10^{3}\, \sigma_{10}^{-4.6}\epsilon_{\star}^{1.38}, 1\right) < n_{0} < 1070 \, \epsilon_{\star}^{0.67} ~,
\end{equation}
and for a single burst
\begin{equation}
    1 < n_{0} < 3.6 \times 10^{3} \, \sigma_{10}^{4.35} N_{\text{SN}}^{-1.3}  ~.
\end{equation}
Comparing the upper and lower limits for the density, we find that the cyclic case requires
\begin{equation}
    \epsilon_{\star} < 0.93 \sigma_{10}^{6.5} ~,
\end{equation}
and a single burst only triggers star formation if
\begin{equation}
    N_{\text{SN}} < 550 \sigma_{10}^{3.35}~.
\end{equation}
Both requirements are readily fulfilled for typical values of the parameters.

\section{Results}\label{sec:results}

\begin{figure*}
\centering
\includegraphics[width=0.95\textwidth]{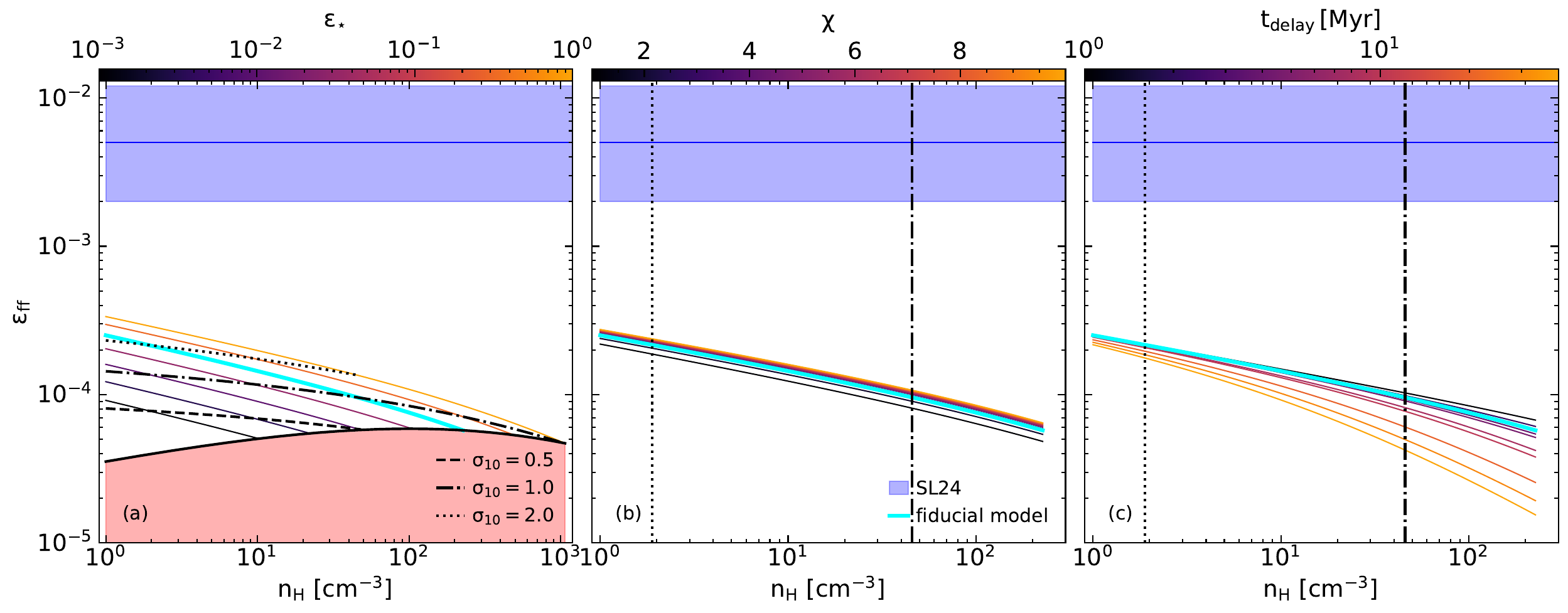}
\caption{Star-formation efficiency per free-fall time as a function of density for various models in the cyclic scenario. In panels (a), (b) and (c) the star-formation efficiency $\epsilon_{\star}$, cloud overdensity $\chi$ and stellar-population lifetime $t_{\text{delay}}$ are respectively varied, while the remaining two parameters are kept at their respective fiducial values of $\left(\epsilon_{\star}^{\text{fid}}, \chi^{\text{fid}}, t_{\text{delay}}^{\text{fid}} \right) = (0.1, 3, 3 \, \text{Myr})$.
Below the solid black line in panel (a) not enough stars are formed to reliably maintain cyclic star-formation (eq. \ref{eq:n_max}). The dashed, dot-dashed and dotted black lines correspond to the lower limit for the density below which shock break-out is expected to affect the dynamics for various values of the ISM velocity dispersion $\sigma_{10}$ (eq. \ref{eq:n_lower_limit}).
The blue line and shaded area correspond to the value of $\epsilon_{\text{ff}}$ and its uncertainty derived from observations on scales of $\sim 100 \, \text{pc}$ \citep{2024arXiv240319843S}.}\label{fig:SFE}
\end{figure*}

In the previous section we have described a model to estimate the star-formation efficiency per free-fall time $\epsilon_{\text{ff}}$ for SN-implosion-triggered star-formation. Here we explore how our model compares to observational estimates of $\epsilon_{\text{ff}}$ and its dependency on the model parameters.

\subsection{Cyclic Star-Formation}

In the cyclic scenario, the number of SNe per cycle is determined by eqs. \ref{eq:N_SN} and \ref{eq:cloud_mass}. In the limit of $t_{\text{GI}} / t_{\text{cycle}}\sim 1$, we thus find
\begin{equation}\label{eq:scaling_cyclic}
    \epsilon_{\text{ff}} \propto \epsilon_{\star}^{0.3} \, n_{0}^{-0.22}\,.
\end{equation}

In Figure \ref{fig:SFE} we show how $\epsilon_{\text{ff}}$ varies with respect to the model parameters $\epsilon_{\star}$, $\chi$ and $t_{\text{delay}}$ and compare it to the most recent observational estimate of $\epsilon_{\text{ff}}$ as reported by \citet{2024arXiv240319843S}.

Panel (a) shows the variation of $\epsilon_{\text{ff}}$ with respect to $\epsilon_{\star}$. The range of densities where cyclic star-formation can be maintained becomes increasingly narrow as $\epsilon_{\star}$ is reduced, as expected from eq. \ref{eq:n_max}.
In media with $\sigma < 45 \, \text{km/s}$ and thus sufficiently small scale-height, shock break-out constrains the range of densities where our model for cyclic star-formation can be applied.
Furthermore, it becomes clear that for fixed density, $\epsilon_{\text{ff}}$ scales as expected from eq. \ref{eq:scaling_cyclic}, since $t_{\text{cf}} \ll t_{\text{GI}} + t_{\text{ff, cloud}}$ for $n_{0} \geq 1$.

Panel (b) shows the variation of $\epsilon_{\text{ff}}$ with respect to $\chi$. Higher $\chi$ generally implies higher $\epsilon_{\text{ff}}$, with a diminishing effect as $\chi \gtrsim 4$. 
The effect diminishes because $t_{\text{GI}} \sim 3\, t_{\text{ff}}$ and both timescales scale similarly with density. Thus, reducing $t_{\text{ff, cloud}} \propto t_{\text{ff}}$ will reduce the importance of $t_{\text{ff, cloud}}$ relative to the much longer $t_{\text{GI}}$, but hardly affect $\epsilon_{\text{ff}}$.

Panel (c) shows the variation of $\epsilon_{\text{ff}}$ with respect to $t_{\text{delay}}$. For $t_{\text{delay}} \ll 10 \, \text{Myr}$ only a slight reduction in $\epsilon_{\text{ff}}$ is noticeable at high ambient densities. 
For longer delay times, comparable to $t_{\text{ff}}$, the effect is stronger.
Since the details of cloud formation are independent of $t_{\text{delay}}$ and only $t_{\text{cycle}}$ depends on it affine-linearly, $\epsilon_{\text{ff}} \propto t_{\text{delay}}^{-1}$ in the limit where $t_{\text{delay}} \gg t_{\text{exp}}$, i.e. high densities or high $t_{\text{delay}}$. 
However, we do not expect this limit to play a big role, as extremely high values of $t_{\text{delay}} \gtrsim 40 \, \text{Myr}$ are unexpected and even at the maximum density eq. \ref{eq:n_max} $t_{\text{delay}} < t_{\text{exp}}$ indicating that the role of $t_{\text{delay}}$ is almost negligible for SN-implosion-triggered star-formation.

The range of values of $\epsilon_{\text{ff}}$ obtained with the fiducial set of parameters is about two orders of magnitude below that derived from observations $\epsilon_{\text{ff}}^{\text{obs}} = 0.5^{+0.7}_{-0.3} \, \%$ with a relatively weak scaling with density $\epsilon_{\text{ff}} \propto n_{0}^{-0.22}$. 
The fact that we cannot explain $\epsilon_{\text{ff}} \sim 0.5\,\%$ is however not a problem and rather a feature as we expect several other processes such as classical triggered star-formation \citep[see e.g.][]{1998ASPC..148..150E} or spontaneous gravitational collapse \citep{1989ApJ...344..685K, 2007ARA&A..45..565M} to contribute to the star formation as well. 

Given that the mass of the central cloud is significantly smaller than the total swept up mass, it is not surprising that the contribution to the star formation is minor. 
However, it should be noted, that the stars formed in the implosion cloud are expected to be more chemically enriched than the bulk of the stars formed by other processes \citep{2024arXiv240205796R}. 
Moreover, we note that the process is self-regulated, i.e. the cloud makes up a larger fraction of the swept up mass for fewer SNe, which motivates us to study the single-burst scenario, which we explore below.

\subsection{Single-Burst Star-Formation}

\begin{figure*}
\centering
\includegraphics[width=0.95\textwidth]{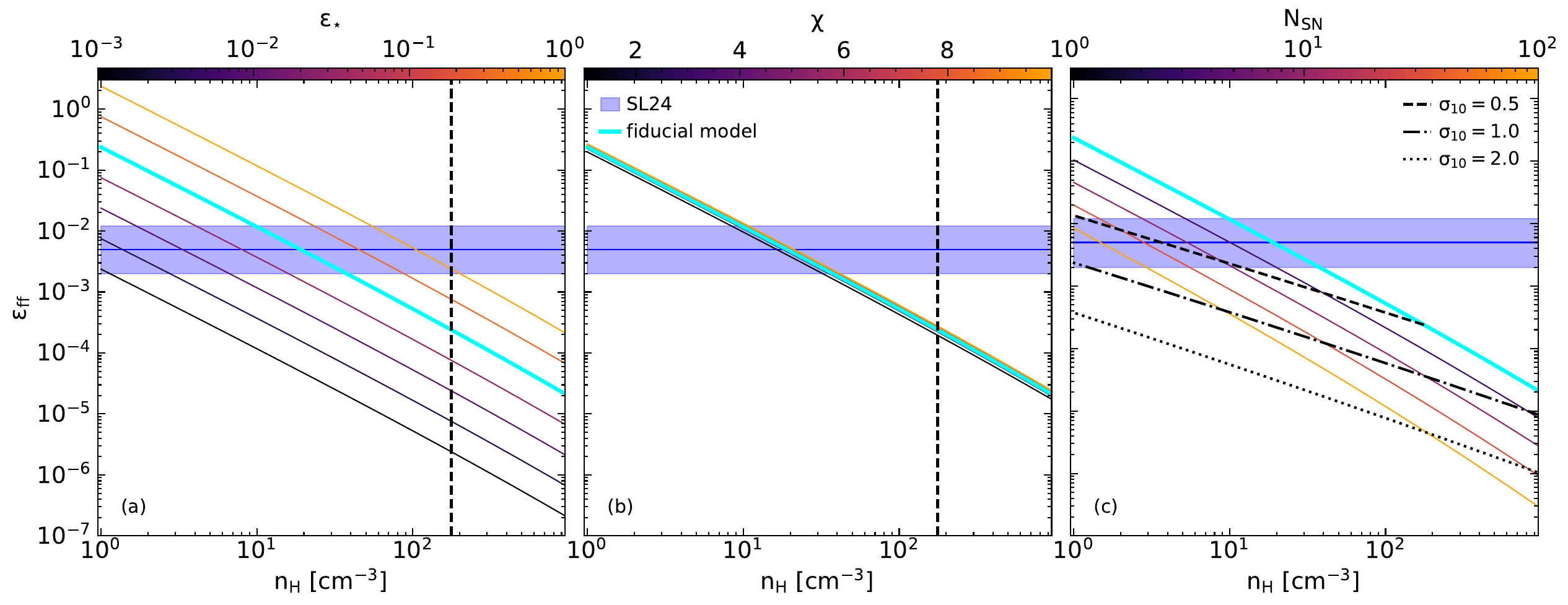}
\caption{Same as Figure \ref{fig:SFE}, but in the case of the single-burst scenario. Panel (c) shows the dependency of $\epsilon_{\text{ff}}$ with respect to $N_{\text{SN}}$, with a fiducial value of $N_{\text{SN}}^{\text{fid}} = 1$. Here the dashed, dot-dashed and dotted black lines correspond to the upper limit of the density above which shock break-out is expected to affect the dynamics for various values of the ISM velocity dispersion $\sigma_{10}$ (eq. \ref{eq:z_sc_constraint}).}\label{fig:SFE_SB}
\end{figure*}

In the single-burst scenario, we consider $\epsilon_{\text{ff}}$ for a single iteration of SN-implosion-triggered star-formation. 
Such an iteration could be triggered by the collective feedback of an entire stellar population or the explosion of a single star, which includes both type-Ia and type-II SNe.
It is thus reasonable to measure $t_{\text{cycle}}$ from the time of the explosion, as the delay time bears no meaning in this case, i.e. setting $t_{\text{delay}} = 0$.

As opposed to the cyclic case, in the single-burst scenario $N_{\text{SN}}$ is not determined by eq. \ref{eq:N_SN}, but instead is treated as a free parameter. 
This implies, that now in the limit of $t_{\text{cycle}} \sim t_{\text{GI}}$, 
\begin{equation}
    \epsilon_{\text{ff}} \propto \epsilon_{\star}\, N_{\text{SN}}^{-0.7}\, n_{0}^{-1.26}\,.
\end{equation}
These scalings are reflected by panels (a) and (c) in Figure \ref{fig:SFE_SB}. Panel (b) shows, that as for the cyclic case, the effect of $\chi$ is small, since $t_{\text{cycle}}$ is dominated by $t_{\text{GI}}$.

Due to the strong scaling with density, the range of values of $\epsilon_{\text{ff}}$ obtained with the fiducial set of parameters spans four orders of magnitude. 
At low densities, star formation is very efficient with $\epsilon_{\text{ff}} \sim 10\,\%$ exceeding the observational mean by over an order of magnitude, while at higher densities it becomes increasingly inefficient.

The fact that $\epsilon_{\text{ff}}$ can significantly exceed the global average derived from observations does, however, not cause any conceptual problems, as the single-burst case should only occur occasionally. 
Thus, while the single-burst scenario might contribute to the scatter, it is not expected to dominate the global average.

\section{Discussion}\label{sec:discussion}

In the previous section we have shown the efficiency of SN-implosion-triggered star-formation in two different scenarios.
Here, we discuss some of the limitations of our model, and the implications of our findings in the broader context of galaxy evolution.

\subsection{Limitations}\label{sec:limitations}

The model considered in this letter is a detailed analysis of the implications of the results of \citet{2024arXiv240205796R}. 
As such, the limitations applying to their results also apply to this model.

Neglecting the turbulent nature of the ISM likely affects both the timescales involved as well as the properties of the central cloud. 
Similarly, as discussed by \citet{2024arXiv240205796R} including more complete physics models for processes, such as magnetic fields, cosmic rays, and heat conduction, could qualitatively modify the cloud-formation process. 
Finally, the inclusion of early stellar feedback and realistic time delays between SNe can further delay the implosion and thus cloud formation \citep[see e.g.][]{2024arXiv240211008F}, leading to a lower star-formation efficiency.

A model is only as good as its assumptions.
It is thus worthwhile to consider the role of our model assumptions.

For sufficiently massive clouds, assuming an IMF seems to be justified by the observed statistics of stellar populations \citep{2021arXiv211210788K}. 
However, for clouds with small masses, the IMF is not well sampled in the high-mass end and the number of massive stars that are formed per cycle will be dominated by stochasticity \citep[see e.g.][]{2023OJAp....6E..48G}.
Nonetheless, since our cyclic model considers a long-term average, populations with $N_{\text{SN}} > 1$ should -- over a larger number of iterations -- sample the high-mass end of the IMF well enough to justify our treatment. 
We have checked eq. \ref{eq:N_SN} for different shapes of the IMF commonly used in the literature \citep[e.g. the list of IMFs in][]{2017AJ....153...85S} and have found only minor differences; The result that there is about 1 SN per $100 \, M_{\odot}$ of newly formed stars is on strong footing.

Clearly it is unreasonable to expect the ambient medium to return \textit{exactly} to its initial state of a uniform medium with constant density at rest. 
Nonetheless, in the context of the simulations of \citet{2024arXiv240205796R} it is revealed that after cloud formation the density in the region between the shock and the cloud indeed approaches the initial density of the ISM as the cavity is refilled with backflowing material from the ever broadening front of the blastwave, while the remaining kinetic energy is dissipated through small-scale shocks, leaving a nearly static, and slightly turbulent medium.

Likewise, we expect the density in the cloud to be locally enhanced after its collapse. However, as long as the SNR can break out of the cloud before shell formation the dynamics are not expected to be strongly influenced. Furthermore, in a more complete picture early feedback might counteract this density enhancement and with more realistic delays between the SNe the central mass distribution is expected to have only a small effect compared to the potentially large time delay before the last SN of the population explodes.

In a more realistic galactic environment, however, these considerations are confounded by several factors. 
The complex interplay of neighboring shocks, gravitational torques, shear,  and stratification effects is expected to lead to significant inhomogeneities in the density and velocity distribution.

We expect the analysis of the single-burst case to be less affected, since the ISM in the immediate vicinity of the SNR is not expected to change dramatically during one cycle.

On the other hand, in the cyclic case, we consider the long-term average, which means that necessarily all of these effects will have enough time to affect the dynamics.
As long as these effects do not prevent cloud formation or disrupt the central clouds before they can collapse, we expect the effect to be small.
Nonetheless, in order to be certain a more thorough analysis is required.

We expect that the effects of vertical stratification can be neglected if the SNR is contained within a scale height of the galactic ISM. By assuming vertical hydrostatic equilibrium in an isothermal, single-component disk we find that this constrains the range of densities for which our model can be applied.
These constraints however scale very strongly with the velocity dispersion and are likely negligible at high redshift, where the velocity dispersion often exceeds 100 km/s \citep{2016MNRAS.458.1671K}.

Moreover, in the regime where the effective turbulent pressure dominates the ambient pressure, we expect the implosion to be launched even earlier due to the increased ambient pressure, reducing the importance of the stratification constraint.
On the other hand, in this turbulent regime we expect larger density fluctuations and more overall anisotropy which would affect our conclusions in a way that is out of the scope of the simple model presented in this work.

Finally, we have neglected the role of other sources of star formation and feedback.
While theoretical considerations suggest that individual SN shocks are an unlikely candidate for triggered star-formation \citep{2011EAS....51...45E}, we anyway consider the potential contribution, given the uncertainty of the subject.

Classical, triggered star-formation has been estimated to be delayed by about a free-fall timescale \citep{1998ASPC..148..150E}.
For low ambient densities, we thus do not expect it to strongly affect the mass budget for cloud- and subsequent star-formation by SN implosion, while at sufficiently high densities, the onset of triggered star-formation might be before the onset of cloud formation and thus reduce the mass budget for cloud formation.

On the other hand, we expect feedback from massive stars formed by classical, triggered star-formation to interfere with the process in several ways.
Feedback that is sourced sufficiently far away from the explosion center might heat the ambient medium and increase its pressure, leading to an earlier implosion and thus accelerate the star-formation process.
However, if the source of the feedback is too close, it might disrupt the cloud before it can form stars and delay or even prevent further star formation, leading to a reduction of the free-fall efficiency.

\subsection{Implications for Galaxy Evolution} \label{sec:implications}

We have shown that SN-implosion-triggered star-formation can contribute $\lesssim 5\,\%$ of the globally averaged star formation. 
While this might not seem like a large contribution, it should be noted that a bulk fraction ($\gtrsim 10\,\%$) of the highly enriched SN ejecta are expected to be locked up in the central, star-forming cloud.
Indeed, if would be problematic if $\epsilon_{\text{ff}}$ were much higher, as this would potentially indicate that our model predicts too much metal-rich star-formation, the implications for which we are discussing in the following.
%Here we discuss the potential implications for metal-rich star-formation.

The increment in metallicity per cycle can be expressed in terms of the fraction of ejecta material locked up in the cloud $M_{\text{ej, cl}} / M_{\text{cl}}$, the metallicity of the ejecta $Z_{\text{ej}}$ and the metallicity of the ambient medium $Z_{0}$:
\begin{equation}
    \delta Z = \frac{M_{\text{ej, cl}}}{M_{\text{cl}}} \left(Z_{\text{ej}} - Z_{0}\right) \,.
\end{equation}

The simulations of \citet{2024arXiv240205796R} suggest that $\gtrsim 10-50\, \%$ of the ejecta end up in the cloud before it collapses.
Indeed, in the absence of numerical diffusion, for a spherically symmetric explosion, one expects $100\, \%$ of the ejecta to be locked up in the central cloud, since radial shells cannot cross in ideal hydrodynamics.

\citet{2019MNRAS.484.2632S} have computed the yields from various feedback processes for a Chabrier IMF with mass range 0.1 to 120 M$_{\odot}$ \citep{2003PASP..115..763C} using \textsc{celib} \citep{2017AJ....153...85S}. 
They find that for metallicities $Z \gtrsim 10^{-6}$, stellar populations eject $\sim 10 \, \%$ of their mass in type-II SNe, with $Z_{\text{ej}} \sim 15 - 20 \, \%$.
In the more uncertain case of primordial metallicity the stellar populations eject $\lesssim 90 \, \%$ of their mass in type-II SNe, with $Z_{\text{ej}} \sim 33\,\%$. 
While their IMF differs slightly from the one we have adopted here, we expect the results to only change slightly.

Combining these results and relating the mass of the stellar population to the cloud mass through $\epsilon_{\star}$ we find that for sufficiently high metallicity $\delta Z \gtrsim 10^{-3} \, \epsilon_{\star}$ and for primordial metallicity $\delta Z \gtrsim 3 \times 10^{-2} \, \epsilon_{\star}$. 

A single cycle takes about $t_{\text{cycle}} \lesssim 10^{7}\,-10^{8}\,\text{yr}$, so over the lifetime of the universe there could have been $\lesssim 100 - 1000$ cycles, during most of which the metallicity would have been non-primordial. 
Neglecting physical dilution effects (but implictly accounting for numerical dilution, which may account for a similar degree of dilution), we thus expect a maximum enrichment of $\Delta Z_{\text{max}} = N_{\text{cycle}} \times \delta Z \sim \left(0.1 - 1\right)\, \epsilon_{\star}$, slightly higher than the metallicities of the most metal-rich stars that have been observed \citep[see e.g.][]{2015ApJ...809..143D, 2024ApJ...960...83S, 2024arXiv240601706R}.

We note that while the enrichment per generation might seem small, it still represents a massive enhancement compared to the expected enrichment one would obtain if the ejecta were instead fully mixed. 
Typically $M_{\text{cl}} / M_{\text{exp}} \sim 10^{-2} - 10^{-3}$, so without confining the ejecta into the central cloud, it would take $10^{4} - 10^{5}$ cycles to reach solar level metallicities and beyond.
Of course, since globally star-formation is $\sim 100$ times more efficient this difference can be overcome and on average solar level metallicities are commonly reached by today.

As the chemical enrichment in this scenario is primarily due to type-II SNe, we expect an elevated level of $\alpha$-elements in the resulting stellar populations.

However, we note that we have neglected other enrichment mechanisms such as type-Ia SNe and stellar winds from asymptotic giant branch and massive stars \citep{1999IAUS..193..670K, 2017AJ....153...85S}, which would lead to even higher metallicities and less extreme $\alpha$-abundances.
On the other hand, dynamical effect such as turbulence and galactic shear could counteract the local self-enrichment by mixing the enriched gas with lower metallicity, ambient gas and limiting the maximum number of cycles during which the system remains coherent. A detailed study of these limitations is out of the scope of our simple model.

In summary, implosion triggered star formation can \textit{locally} accelerate chemical evolution. If left alone, the thus formed stellar populations can reach super-solar metallicities long before it would be possible for populations that are formed in clouds that were mainly enriched by \textit{external} sources. Ultimately, the maximum metallicities and the number of highly enriched stars that can be reached through implosion triggered star formation depend on the timespan during which the system can coherently undergo this process.

\section{Concluding Remarks}\label{sec:summary}

We have analyzed the efficiency of SN-implosion-triggered star-formation and its contribution to the global star formation.
Our analysis reveals that if maintained, star-formation from SN implosion is quite inefficient, contributing only about $\lesssim 5\,\%$ of the observed star-formation efficiency per free-fall time $\epsilon_{\text{ff}}$. 
Nonetheless, because of the projected high metal enrichment of the thus formed stars, we expect this process to contribute significantly to the formation of metal-rich stars.

While our idealized model is useful for obtaining a rough estimate of the contribution from SN-implosion-triggered star-formation, more detailed models are required to explore the role of the galactic environment and external sources of feedback.

We conclude that SN-implosion-triggered star-formation offers a compelling, well-motivated pathway to the formation of metal-rich stars. 

%% Also note that the acknowlodgment environment does not support long amounts of text. If you have a lot of people and institutions to acknowledge, do not use this command. Instead, create a new \section{Acknowledgments}.
\begin{acknowledgments}
We thank the anonymous referee for their insightful comments and suggestions that helped to improve the quality of this work.
This research was funded by the Deutsche Forschungsgemeinschaft (DFG, German Research Foundation) under Germany's Excellence Strategy – EXC 2094 – 390783311.
\end{acknowledgments}

\vspace{5mm}
%\facilities{HST(STIS), Swift(XRT and UVOT), AAVSO, CTIO:1.3m,


\software{\textsc{Julia} v1.6.5 \citep{2014arXiv1411.1607B},
          \textsc{Matplotlib} v3.5.1 \citep{thomas_a_caswell_2021_5773480}
          }



%\appendix


\bibliography{bibliography}{}
\bibliographystyle{aasjournal}



\end{document}